\documentclass{aa} 
\usepackage{graphicx}
\begin{document}

   \title{The Open Cluster Distance Scale}

   \subtitle{A New Empirical Approach}

   \author{Susan M. Percival\inst{1}, Maurizio Salaris\inst{1}
          \and
          David Kilkenny\inst{2}
          }

   \offprints{Susan M. Percival}

   \institute{Astrophysics Research Institute, Liverpool John Moores University,
              Twelve Quays House, Egerton Wharf, Birkenhead, CH41 1LD, UK\\
              \email{smp,ms@astro.livjm.ac.uk}
              \and
             South African Astronomical Observatory, P.O. Box 9, Observatory 7935,
             South Africa\\
             \email{dmk@saao.ac.za} 
             }

   \date{Received ; accepted}

   \abstract{
             We present new $BV(RI)_{\rm C}$ photometry for a sample of 54 local G and K stars with 
accurate \emph{Hipparcos} parallaxes in the metallicity range $-0.4 \leq {\rm [Fe/H]} \leq +0.3$.
We use this sample to develop a completely model-independent main sequence (MS) fitting method which
we apply to 4 open clusters -- the Hyades, Praesepe, the Pleiades and NGC 2516 -- which all 
have direct \emph{Hipparcos} parallax distance determinations.  Comparison of our MS-fitting results
with distances derived from \emph{Hipparcos} parallaxes enables us to explore whether the 
discrepancy between the \emph{Hipparcos} distance scale and other MS-fitting methods found for some
clusters is a consequence of model assumptions.
We find good agreement between our results and the \emph{Hipparcos} ones for the Hyades and Praesepe. 
For the Pleiades and NGC~2516, when adopting the solar abundance determined from spectroscopy, we 
find significant disagreement at a level similar to that found by other MS-fitting studies. 
However, the colour-colour relationship for both these clusters suggests that their metallicity is
significantly subsolar.  Since the MS-fitting method relies on matching the cluster colours to a 
template MS, we argue that, when applying this method, the appropriate metallicity to adopt is the
photometric subsolar one, not the solar abundance indicated by spectroscopy.
Adopting photometric metallicities for all 4 clusters, we find complete agreement with the 
\emph{Hipparcos} results and hence we conclude that the mismatch between the spectroscopic and 
photometric abundances for the Pleiades and NGC~2516 is responsible for the discrepancies in distance
estimates found by previous studies.  The origin of this mismatch in abundance scales remains an 
unsolved problem and some possible causes are discussed.
   }
\authorrunning{Percival et al.}
\titlerunning{Open Cluster Distance Scale}
   \maketitle

\section{Introduction}

The results of the \emph{Hipparcos} astrometric mission (European Space Agency \cite{ESA}) have
recently provided the distances to several open clusters derived from their directly measured 
trigonometric parallaxes. These distances have, in turn, been used to test theoretical stellar 
models and investigate the evolutionary status of open cluster stars, by comparing theoretical
isochrones with cluster Colour-Magnitude-Diagrams (CMDs) (see e.g., Castellani et al.
\cite{castel} and references therein).

It was very soon realized that the distances obtained by fitting the Main
Sequence (MS) of isochrones with the appropriate metallicity to the
corresponding cluster sequence (MS-fitting) do not always agree with the
\emph{Hipparcos} parallax distances.  This was first shown by Pinsonneault
et al.~(\cite{pinson}, hereafter P98), whose theoretical isochrones, and their isochrones
calibrated on the Hyades shape, provide MS-fitting distances in agreement,
within the associated 1$\sigma$ errors, with \emph{Hipparcos} results for
the cases of Hyades, Praesepe and $\alpha$-Persei.  However the Pleiades
MS-fitting result is significantly longer than the \emph{Hipparcos} value,
whilst that of Coma Ber is significantly shorter. The case of the Pleiades
is particularly egregious, with a discrepancy of almost 0.3 mag in the
distance modulus.

Subsequent work has been devoted to the comparison of the Pleiades CMD
with local MS stars of comparable metallicity, and with accurate
parallaxes, in order to avoid the use of any theoretical input in deriving
the MS-fitting distance. Soderblom et al.~(\cite{soder}) selected a sample of
`young' (on the basis of their chromospheric activity) solar-type stars
(therefore supposedly of solar metallicity, like the Pleiades), which
should represent the local counterpart of Pleiades stars; they have hence
shown that the location of this local sample is compatible with the longer
MS-fitting distance, not with the \emph{Hipparcos} one. Only supposedly
more metal poor objects (based mainly on their space velocity, not on
spectroscopic estimates) show a MS location in agreement with the
\emph{Hipparcos} distance.

Stello \& Nissen~(\cite{stello}) considered a sample of local F-stars with
accurate parallaxes and Str\"omgren photometry, and selected objects
within $\sim \pm$ 0.10 of the cluster [Fe/H] -- the Str\"omgren
photometry for both Pleiades and field stars being used to derive the
relative [Fe/H] and reddening values. The observed magnitude
distribution of these field stars at a fixed colour shows a large spread
due to the convolution of observational errors, binary contamination and
evolutionary distribution, since F-stars 1-2 Gyr old already depart from 
the Zero Age MS (ZAMS).  However, the authors were able to determine
statistically a ZAMS locus from the lower envelope of the MS
distribution which, again, was shown to lie above the Pleiades MS if the
\emph{Hipparcos} distance is assumed for the cluster.  For example, in the
$M_{\mathrm V}/(b-y)_0$ plane the discrepancy is $\sim$0.25 mag.

On the other hand, very recently Castellani et al.~(\cite{castel}) have been
able to achieve a good fit of the Pleiades MS with their theoretical
isochrones and the \emph{Hipparcos} distance, employing a subsolar
abundance, namely ${\rm [Fe/H]} \sim -0.15$, but one which is still compatible with
recent metallicity determinations. At the same time their models can fit
the Hyades MS with the \emph{Hipparcos} distance, unlike the case of
P98 whose models reproduce the Hyades distance, but not that of the Pleiades. 
The difference between these 2 theoretical results may be due to different 
sensitivities of the isochrone colours to [Fe/H]; in fact, according to P98 
one needs ${\rm [Fe/H]} < -0.20$ to solve the discrepancy.

These conflicting results about the open cluster \emph{Hipparcos} and MS-fitting distance scales 
are disturbing for reasons with far-reaching consequences, beyond merely the testing of stellar 
evolution models.  Various young clusters harbour Cepheid stars, whilst intermediate-age and
old clusters have Red Clump stars, both classes of objects being widely employed as extragalactic 
distance indicators.  Testing of the properties of these standard candles in order to determine 
accurate zero-point calibrations necessarily makes use of MS-fitting distances to their parent 
cluster, since they are located at distances too large to have accurate \emph{Hipparcos} 
parallaxes.  In addition, MS-fitting distances are used to tie the Galactic globular cluster
distances (and hence the age of the Galaxy) to the \emph{Hipparcos} parallaxes, due to the lack of a 
precise parallax calibration for the RR~Lyrae star absolute brightness (Groenewegen \& Salaris 
\cite{groen}, but see also recent parallax determination for RR Lyrae itself by Benedict et 
al.~\cite{bene}). It is therefore paramount to 
assess the accuracy of MS-fitting distances based on field stars with accurate \emph{Hipparcos} 
parallaxes, by using open clusters on which both techniques can be applied.

In this paper we further analyze the consistency between MS-fitting and direct \emph{Hipparcos} 
parallax distances for the Hyades, Praesepe, Pleiades and
NGC2516 open clusters; these clusters have deep and accurate enough
CMDs to apply with confidence the MS-fitting method.
We employ a purely empirical approach, based on selecting a sample of 54 G-stars from the
local field MS population which we use to construct a template MS, and for 
which we have secured homogeneous $BV(RI)_{\rm C}$ photometry.  Strict selection criteria 
ensure that these stars are not binary, have accurate \emph{Hipparcos} parallaxes, 
and populate a region of the lower MS completely unaffected by age differences between 
field and cluster stars.

The layout of the paper is as follows:  in Section 2 we present the new
field dwarf sample, detailing selection criteria and sample biases, Sect. 3
describes our method for determining [Fe/H] values for the sample stars,
whilst Sect. 4 gives full details of our empirical MS-fitting method.  In Sect. 5 
we present our results for the Hyades, Praesepe, Pleiades and
NGC~2516 and in Sect. 6 we summarize these results and discuss their implications.


\section{The Field Dwarf Sample}

\subsection{Selection Criteria}

Suitable field dwarfs were selected using the three basic criteria of parallax error, absolute
magnitude and metallicity, as described below.  

All the stars in the sample have \emph{Hipparcos} parallaxes with errors $< 12\%$, enabling both
an accurate determination of their absolute magnitudes and an assessment of the Lutz-Kelker bias
in the sample (see Sect. 2.3).  

Absolute magnitudes are limited to stars with $M_{\rm V} \geq 5.4$ to 
ensure they are unevolved and on the lower MS, and hence their position in a CMD does not
depend on their age, which is not known.

Metallicities for the field dwarfs are determined from Str\"{o}mgren indices available in the 
literature (Hauck \& Mermilliod \cite{hauck}).  During the process of sample selection, the 
calibration of Schuster \& Nissen (\cite{sn}) (hereafter SN89) was used to calculate the 
field dwarf metallicities, and the range was limited to $-0.3 \leq {\rm [Fe/H]} \leq +0.3$ to be 
compatible with the range of abundances of the open clusters to be studied.  However, in the early 
stages of this work it became apparent that the SN89 calibration for G dwarfs may not be reliable in 
this metallicity range.  Twarog et al. (\cite{twarog}) find that the SN89 
calibration systematically underestimates abundances for stars with ${\rm [Fe/H]} > \sim -0.2$, and 
that the effect is strongly colour dependent.  The issue of determining metallicities from the 
Str\"{o}mgren indices will be addressed in Sect. 3.

The \emph{Hipparcos} catalogue entries for all suitable candidates meeting the initial criteria
were carefully checked to avoid the inclusion of any binaries or variable stars. 
From this selection process 54 suitable field dwarfs were identified and we have acquired new 
photometric $BV(RI)_{\rm C}$ data for all of them.

\subsection{Observations and Data Reduction}

Johnson-Cousins $BV(RI)_{\rm C}$ data for 54 field dwarfs were obtained at the
Sutherland site of the South African Astronomical Observatory (SAAO)
in 2002 February, March and June using the St Andrews Photometer on the 
1m telescope and, for a few of the brightest stars, the Modular Photometer 
on the 0.5m telescope. Both photometers are of relatively conventional 
design and employ Hamamatsu R943-02 GaAs photomultipliers. All reductions 
were carried out using standard SAAO software and the data were transformed 
to the Cousins E-region system (see Menzies et al. \cite{menzies} for a compilation of 
standard stars). Four stars were observed on both telescopes and used to 
check for zero point differences; these were found to be $\leq$0.002 (and 
therefore insignificant) for all quantities except $(V-I)_{\rm C}$ where a mean 
difference of 0.005 $\pm$ 0.005 was found. This is hardly significant,
but a correction of $-$0.005 was applied to the $(V-I)_{\rm C}$ data from the 0.5m
system to make the total data set as uniform as possible. The mean standard 
deviations of the data sample are $\pm$0.007, 0.004, 0.003 and 0.005 for $V$, 
$(B-V)$, $(V-R)_{\rm C}$ and $(V-I)_{\rm C}$ respectively (126 observations of 54 stars) and we 
do not believe that any residual systematic errors in the data (arising from
transformation errors) would be significantly larger than these values.

Table~\ref{SDdata} presents the new data and lists \emph{Hipparcos} number, observed 
$V$-magnitude, $(B-V)$, $(V-R)_{\rm C}$ and $(V-I)_{\rm C}$ colours, standard deviations (in $m$mags), 
number of observations, parallax and parallax error (in $mas$) and [Fe/H] (calculated as described 
in Sect. 3).  Our empirical MS-fitting method employs the $(B-V)$ and $(V-I)_{\rm C}$ colour indices -- the 
$(V-R)_{\rm C}$ observations are listed for information only.


\begin{table*} 
\renewcommand{\arraystretch}{1.1} 
\caption[]{New Photometry for Field Dwarf Sample}
\begin{minipage}{\textwidth} 
\begin{tabular}{rrcccrrrrcccc} \hline 
            HIP
            & \multicolumn{1}{c}{$V$} 
            & $(B-V)$
            & $(V-R)_{\rm C}$
            & $(V-I)_{\rm C}$
            & \multicolumn{1}{c}{$\sigma_{V}$}
            & \multicolumn{1}{c}{$\sigma_{BV}$}
            & \multicolumn{1}{c}{$\sigma_{VR}$}
            & \multicolumn{1}{c}{$\sigma_{VI}$}
            & ~~~N
            & $\pi$
            & $\Delta\pi$
            & [Fe/H] \\
            \hline
  39088 &  9.239 & 0.824 &  0.449 & 0.843 &  9 &  0 &  1 &  2 &  ~~~2 &  19.52 &  0.83 &   0.334 \\
  39342 &  7.166 & 0.889 &  0.480 & 0.896 &  2 &  3 &  5 &  4 &  ~~~2 &  57.88 &  0.58 &  -0.043 \\
  40051 &  8.778 & 0.897 &  0.502 & 0.937 &  6 &  3 &  3 &  4 &  ~~~4 &  29.86 &  0.82 &   0.090 \\
  40419 &  8.272 & 0.737 &  0.406 & 0.800 &  2 &  5 &  2 &  2 &  ~~~4 &  29.39 &  1.14 &  -0.483 \\
  42074 &  7.330 & 0.819 &  0.445 & 0.841 &  0 &  5 &  0 &  2 &  ~~~2 &  45.95 &  1.01 &   0.044 \\
  42281 &  8.688 & 0.883 &  0.484 & 0.901 &  1 &  5 &  3 &  3 &  ~~~4 &  27.02 &  1.18 &   0.310 \\
  42914 &  8.183 & 0.763 &  0.420 & 0.804 &  5 &  1 &  5 &  1 &  ~~~4 &  32.14 &  0.82 &  -0.095 \\
  44341 &  8.028 & 0.819 &  0.429 & 0.819 &  2 &  2 &  2 &  2 &  ~~~2 &  32.18 &  1.09 &   0.210 \\
  44719 &  8.410 & 0.772 &  0.421 & 0.813 &  4 &  4 &  2 &  7 &  ~~~2 &  25.83 &  0.91 &   0.034 \\
  46580 &  7.203 & 1.020 &  0.577 & 1.065 &  3 &  2 &  1 &  9 &  ~~~2 &  78.87 &  1.02 &  -0.110 \\
  46422 &  8.855 & 0.843 &  0.465 & 0.893 &  5 &  5 &  2 &  3 &  ~~~2 &  25.40 &  0.83 &  -0.204 \\
  48754 &  8.524 & 0.748 &  0.405 & 0.783 &  2 & 10 &  5 & 11 &  ~~~3 &  27.18 &  1.10 &  -0.321 \\
  50032 &  9.068 & 0.859 &  0.470 & 0.884 &  5 &  3 &  1 &  8 &  ~~~3 &  23.19 &  1.09 &   0.022 \\
  50274 &  8.966 & 0.790 &  0.428 & 0.827 &  4 &  3 &  0 &  2 &  ~~~2 &  22.42 &  0.82 &  -0.246 \\
  50713 &  9.360 & 0.781 &  0.423 & 0.807 & 11 &  3 &  1 &  4 &  ~~~3 &  17.30 &  1.26 &   0.134 \\
  50782 &  7.769 & 0.784 &  0.423 & 0.806 & 16 &  6 &  6 &  7 &  ~~~2 &  37.30 &  1.31 &   0.063 \\
  51297 &  8.859 & 0.818 &  0.449 & 0.861 &  4 &  2 &  1 &  3 &  ~~~3 &  29.83 &  1.03 &  -0.318 \\
  54538 &  9.738 & 0.785 &  0.437 & 0.845 &  3 &  2 &  4 &  8 &  ~~~3 &  16.39 &  1.24 &  -0.156 \\
  55210 &  7.275 & 0.747 &  0.413 & 0.788 &  7 &  4 &  1 &  1 &  ~~~2 &  45.48 &  1.00 &  -0.222 \\
  57321 &  9.370 & 0.783 &  0.429 & 0.814 &  0 &  5 &  3 &  3 &  ~~~2 &  18.94 &  1.33 &  -0.073 \\
  58536 &  8.411 & 0.760 &  0.416 & 0.793 &  2 &  1 &  4 &  2 &  ~~~2 &  27.81 &  1.05 &   0.033 \\
  58949 &  8.166 & 0.745 &  0.420 & 0.809 &  3 &  2 &  0 &  0 &  ~~~2 &  30.58 &  1.02 &   0.008 \\
  59572 &  7.918 & 0.800 &  0.426 & 0.791 &  3 &  6 &  2 &  0 &  ~~~2 &  32.30 &  1.02 &   0.374 \\
  59639 &  8.633 & 0.902 &  0.499 & 0.932 &  7 &  5 &  0 &  1 &  ~~~2 &  31.54 &  0.83 &   0.027 \\
  61291 &  7.143 & 0.848 &  0.480 & 0.896 &  0 &  4 &  0 &  0 &  ~~~2 &  61.83 &  0.63 &  -0.205 \\
  61998 &  8.441 & 0.726 &  0.403 & 0.770 &  3 &  2 &  1 &  2 &  ~~~2 &  27.45 &  1.13 &  -0.242 \\
  62942 &  8.247 & 0.861 &  0.475 & 0.903 &  8 &  2 &  0 &  3 &  ~~~2 &  38.12 &  1.44 &  -0.094 \\
  64103 &  9.668 & 0.696 &  0.385 & 0.746 &  7 &  3 &  6 &  8 &  ~~~2 &  14.41 &  1.48 &  -0.290 \\
  64125 &  9.402 & 0.816 &  0.459 & 0.878 &  6 &  0 &  1 &  0 &  ~~~2 &  19.08 &  1.28 &  -0.288 \\
  65121 &  8.594 & 0.949 &  0.524 & 0.969 &  2 &  2 &  2 &  1 &  ~~~2 &  32.83 &  1.08 &   0.142 \\
  65646 & 10.773 & 0.998 &  0.599 & 1.112 &  4 &  6 &  0 &  1 &  ~~~2 &  18.90 &  2.38 &  -0.401 \\
  67344 &  8.340 & 0.829 &  0.450 & 0.848 &  9 &  3 &  2 &  2 &  ~~~2 &  31.78 &  1.06 &   0.052 \\
  68936 &  8.356 & 0.833 &  0.443 & 0.824 &  0 &  7 &  1 &  5 &  ~~~2 &  26.14 &  1.18 &   0.444 \\
  69075 &  9.495 & 0.956 &  0.559 & 1.040 &  2 &  0 &  1 &  1 &  ~~~2 &  28.79 &  1.36 &  -0.490 \\
  69301 & 10.760 & 0.905 &  0.500 & 0.929 &  1 &  0 &  0 &  2 &  ~~~2 &  15.21 &  2.44 &  -0.208 \\
  69357 &  7.938 & 0.869 &  0.480 & 0.906 & 10 &  3 &  1 &  2 &  ~~~3 &  43.35 &  1.40 &  -0.081 \\
  69570 &  8.235 & 0.673 &  0.375 & 0.732 &  4 &  2 &  0 &  3 &  ~~~2 &  27.62 &  1.12 &  -0.557 \\
  71673 & 10.203 & 0.770 &  0.430 & 0.827 &  3 &  0 &  0 &  2 &  ~~~2 &  16.33 &  1.62 &  -0.182 \\
  72312 &  7.761 & 0.896 &  0.500 & 0.935 &  1 &  0 &  1 &  1 &  ~~~2 &  50.84 &  1.04 &  -0.123 \\
  72339 &  8.046 & 0.786 &  0.423 & 0.804 &  9 &  2 &  2 &  9 &  ~~~2 &  33.60 &  1.51 &   0.133 \\
  72577 &  9.073 & 0.972 &  0.553 & 1.030 &  9 &  2 &  0 &  7 &  ~~~2 &  32.53 &  1.56 &  -0.301 \\
  72688 &  7.804 & 1.014 &  0.583 & 1.069 &  2 &  4 &  0 &  3 &  ~~~2 &  58.96 &  1.05 &  -0.027 \\
  72703 &  8.380 & 0.702 &  0.381 & 0.746 &  4 &  2 &  0 &  6 &  ~~~2 &  25.68 &  1.29 &  -0.386 \\
  73547 &  7.736 & 0.719 &  0.397 & 0.768 &  0 &  4 &  1 &  4 &  ~~~2 &  36.84 &  0.98 &  -0.522 \\
  73963 & 10.343 & 0.898 &  0.503 & 0.943 & 10 &  2 &  5 &  6 &  ~~~3 &  13.57 &  1.91 &  -0.116 \\
  75266 &  8.281 & 0.993 &  0.542 & 0.990 & 12 &  2 &  1 &  5 &  ~~~3 &  39.35 &  1.37 &   0.155 \\
  80043 &  8.901 & 0.952 &  0.550 & 1.032 &  0 &  0 &  1 &  0 &  ~~~2 &  38.80 &  1.37 &  -0.456 \\
  80700 &  8.807 & 0.779 &  0.422 & 0.803 &  6 &  1 &  1 &  3 &  ~~~3 &  21.50 &  1.27 &   0.314 \\
  81237 &  8.756 & 0.773 &  0.426 & 0.812 & 13 &  0 &  1 &  3 &  ~~~3 &  25.32 &  1.15 &  -0.149 \\
  84164 &  9.186 & 0.852 &  0.504 & 0.953 &  2 &  2 &  0 &  4 &  ~~~2 &  17.74 &  1.51 &  -0.304 \\
  85425 &  7.879 & 0.692 &  0.390 & 0.745 &  1 &  2 &  2 &  0 &  ~~~2 &  32.04 &  1.24 &  -0.432 \\
  87089 &  8.912 & 0.823 &  0.474 & 0.898 & 10 &  9 &  9 & 12 &  ~~~2 &  26.16 &  1.16 &  -0.112 \\
  88553 &  8.460 & 0.718 &  0.403 & 0.766 &  3 &  0 &  0 &  2 &  ~~~2 &  26.99 &  1.19 &  -0.149 \\
  89497 &  8.545 & 0.760 &  0.421 & 0.806 &  9 &  3 &  2 &  0 &  ~~~2 &  27.11 &  1.20 &  -0.088 \\
\hline 				
\label{SDdata} 
\end{tabular} 
\vspace*{-0.6cm} 
\end{minipage} 
\end{table*}

\subsection{Lutz-Kelker Corrections, Reddening and Metallicity Bias}

Since the field dwarf sample was selected on parallax error, their absolute magnitudes are subject 
to Lutz-Kelker bias which has the effect of systematically underestimating their distances 
(Lutz \& Kelker \cite{lk}).  Following the procedures of Hanson (\cite{hanson}), Lutz-Kelker 
corrections were derived based on the distribution of proper motions of the sample stars, which 
were taken from the \emph{Hipparcos} database.  Hence, absolute magnitudes were corrected using
the relation:
$$ 
\Delta M_{\rm LK} = -7.60(\sigma_{\pi}/\pi)^{2} - 47.20(\sigma_{\pi}/\pi)^{4}
$$ 
Individual LK corrections are very small since 43 out of the 54 stars have parallax errors 
$< 5 \%$ and in fact the average correction is less than 0.02 mag.

Reddening for the field dwarfs was neglected since all the stars in the sample lie well within 75 pc 
of the Sun (the average distance being 35.9 pc), a region in which several authors have concluded
that reddening effects are negligible (see discussion in Percival et al. \cite{percival}, 
hereafter Paper I).

We made no attempt to correct for any metallicity bias in the sample since our previous
investigation showed that any effect on the derived cluster distances is very small, even when 
working in a lower metallicity regime (see discussion in Paper I).  For the current sample, the 
peak of the metallicity distribution falls roughly between ${\rm [Fe/H]} = -0.2$ and solar, which 
reflects the distribution of the parent population in the solar neighbourhood and so any biases 
are likely to be negligible.  


\section{Metallicity Calibration}

The MS-fitting method relies crucially on determining, on an homogeneous scale, the metallicities 
of a) the clusters being studied and b) the field dwarf sample, which is used to construct a template
main sequence.  In our earlier 
study of the Galactic Globular Cluster 47 Tucanae (Paper I), we used Str\"{o}mgren data and the 
G-star calibration of SN89 to calculate abundances for a sample of local subdwarfs in the range
$-1.0 < {\rm [Fe/H]} < -0.3$ -- a procedure adopted by many other authors for deriving abundances of 
sub-solar metallicity stars.  However, Twarog et al. (\cite{twarog}) find a discrepancy in this 
G-star calibration which 
came to light when analyzing cool dwarfs of late spectral type in the Hyades, for which the 
abundance is known to be super-solar.  The effect appears to be systematic and strongly colour 
dependent and can lead to an underestimate of the abundance of solar or super-solar abundance stars
by as much as 0.4 dex.  This problem may begin to affect stars with abundances greater than
${\rm [Fe/H]} \sim -0.2$ and is most pronounced when $b-y$ is larger than $\sim$ 0.47.  Since in our 
study we are using exclusively unevolved, lower MS stars at roughly solar metallicity, these are 
the very stars which are most strongly affected.

Twarog et al. (\cite{twarog}) suggest that at least one of the sources of the problem lies in 
the metallicity dependence
of the Str\"{o}mgren $c_{1}$ index, which has previously been underestimated for more metal-rich 
stars.  This results in some main-sequence stars with high metallicities (above the Hyades 
value) being classified as giants because of their high $c_{1}$ index, which in turn can lead
to systematic discrepancies when these indices are used to derive abundances.  The original
SN89 calibration was intended to apply across a large metallicity range, 
$-2.6 < {\rm [Fe/H]} < 0.4$, and accounted for the fact that some stars have started to evolve off the
main sequence.  Hence, it has a complicated functional form with many cross terms, making it
virtually impossible to disentangle the metallicity dependence of one particular term 
($c_{1}$ in this case).
In an attempt to address this problem, Haywood (\cite{haywood}, hereafter H02) has provided an 
updated version of the SN89 calibration, using more recent spectroscopic data for the calibrating 
stars, which keeps the same functional form and redetermines the coefficients of the various terms.
Again, this calibration is valid over a wide metallicity range, $-2.0 < {\rm [Fe/H]} < 0.5$, and
accounts for evolving stars.  Both the SN89 and the H02 calibrations require
the metallicity of the Hyades to be specified as an input parameter, specifically, Hyades stars 
are included in the fitting procedure with an assumed Hyades metallicity, and given double 
weighting to force the fit at the high metallicity end of the scale.  SN89 assumed
${\rm [Fe/H]_{Hyades}}=0.13$, whilst H02 adopts a value of 0.14.

Since all the field dwarfs in our MS-fitting sample are specifically chosen to be unevolved stars
(ensured by the requirement that $M_{\rm V} \geq 5.4$) and should all be within $\sim$ 0.4 dex of
solar metallicity (specifically, they are not low metallicity halo stars, but rather local disk
stars) the metallicity calibration from Str\"{o}mgren indices can potentially be a lot simpler.
The most important element in any attempted metallicity calibration is to have an homogeneous
sample of spectroscopically determined abundances for stars for which there are also Str\"{o}mgren
data.  It is also vital to be able to tie the abundance scale of any particular sample to some
known zero-point, so that useful comparisons can be made.  

For this purpose, we chose to use the sample of Favata et al. (\cite{favata}) (hereafter FMS97), 
who present [Fe/H] determinations from homogeneous spectroscopic data for a volume limited 
sample of G and K dwarfs from the Gliese catalogue of nearby stars.  FMS97 compare their abundance 
scale with the catalogue of Taylor (\cite{taylor2}, and references therein), in which literature 
[Fe/H] values for cool stars are converted to a coherent temperature scale (the same temperature 
scale as that adopted by FMS97).  They find almost one-to-one correspondence for the stars in common 
between the 2 samples, the slope of the mean relationship being 1.028 with a RMS scatter of 0.09,
and negligible offset (see FMS97, their figure 2). Taylor (\cite{taylor1}) finds 
${\rm [Fe/H]} = 0.107 \pm0.01$ for the Hyades, a value which is consistent with the scale of Gratton 
(\cite{gratton}, hereafter G00) for open clusters, which we will use in this work.  G00
consists of a self-consistent set of abundance determinations for open clusters compiled from
values in the literature, and calibrated against high resolution spectroscopic data, the listed 
value for the Hyades being ${\rm [Fe/H]} = 0.13 \pm 0.06$.
Hence we adopt the metallicity scale of FMS97 for our calibrating sample without any change in
zero-point.

We identified all the stars in the FMS97 sample which met the same basic criteria as our field dwarf 
sample and had Str\"{o}mgren data in the Hauck \& Mermilliod (\cite{hauck}) catalogue.  Hence a cut
in magnitude was made at $M_{\rm V} \geq 5.4$ (derived from \emph{Hipparcos} parallaxes and photometry 
in the literature) to ensure that the calibrating stars were unevolved, and the metallicity range 
was limited to $-0.45 \leq {\rm [Fe/H]} \leq +0.35$.
This resulted in a subset of 18 stars.  We then took all the Hyades members listed as single and
non-variable in Perryman et al. (\cite{perry}) with Str\"{o}mgren indices in the Hauck \& 
Mermilliod catalogue and made a cut
in $b-y$ of $0.31 \leq b-y \leq 0.79$;  this corresponds roughly to an absolute magnitude cut 
of $3.5 < M_{\rm V} < 8.0$ and yielded a subset of 41 stars.  We then examined the relationship 
between the Str\"{o}mgren $b-y$, $c_{1}$ and $m_{1}$ indices and the spectroscopic metallicities for 
the FMS97 sample, by comparing them with the Hyades main sequence relationships.  As suggested 
by Twarog et al. (\cite{twarog}), we found that the principal metallicity indicator for these 
unevolved stars is the $c_{1}$ index.  We then proceeded as follows:  

The Hyades main sequence in $c_{1}$ vs. $m_{1}$ was fitted with a 4th order polynomial via a 
least-squares fitting routine.  The best-fit relationship was found to be:
\newline 
$c_{1_{\rm Hyades}}~=~12.3608m_{1}^{4}-25.5504m_{1}^{3}+18.00m_{1}^{2}$

\hskip2.5cm
$-5.4054m_{1}+0.9067$

\begin{figure}
     \resizebox{\hsize}{!}{\includegraphics{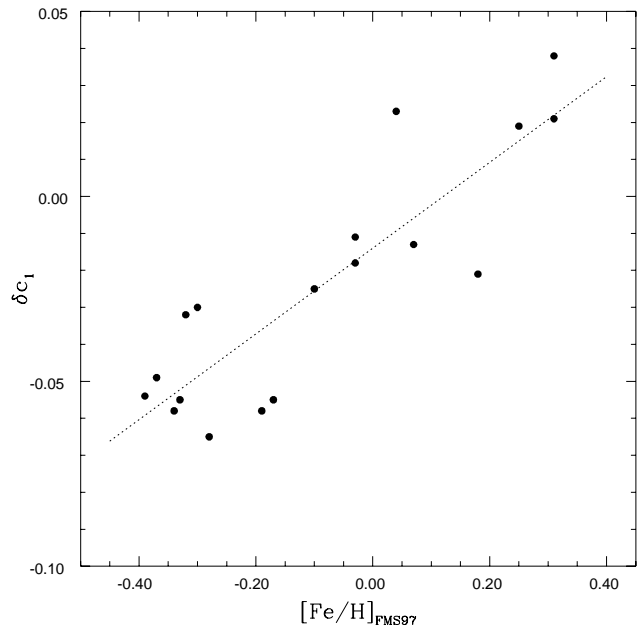}}
      \caption{$\delta c_{1}$ as function of metallicity for the FMS97 subset used in 
               our calibration.  The dotted line shows the line of best-fit.
              }
         \label{dc1fit}
   \end{figure}

Then, for each of the stars in the FMS97 subset, at its value of $m_{1}$, the measured $c_{1}$ was 
compared with the predicted $c_{1}$ from the Hyades relation above, to yield $\delta c_{1}$, where 
$\delta c_{1} = c_{1} - c_{1_{\rm Hyades}}$.
$\delta c_{1}$ was then plotted as a function of the FMS97 metallicity, and the best-fit relation was 
found to be:
$$
\delta c_{1} = 0.116{\rm [Fe/H]_{FMS97}} - 0.014
$$ 
(see Fig.~\ref{dc1fit}).


Applying these relationships to the individual stars in the FMS97 sample to `predict' their 
metallicities, and comparing these values with their spectroscopic determinations, we find that
$$
{\rm [Fe/H]_{predicted}} = 1.004{\rm [Fe/H]_{FMS97}} + 0.004
$$ 
with a dispersion of 0.14 dex.  Applying the same procedures to the 41 Hyads yields an average
metallicity of ${\rm [Fe/H]}=0.122$, also with a dispersion of 0.14 dex.  Is it important to note that
this calibration \emph{does not assume} a known metallicity for the Hyades, but rather, predicts
it (on the FMS97 scale).

A comparison of our predicted metallicities with those of H02 is rather good -- for our full
field dwarf sample, and including the 41 Hyads used in our calibration, the line of best-fit is 
${\rm [Fe/H]_{us}} = 0.946{\rm [Fe/H]_{Haywood}} + 0.006$, with a dispersion of 0.13 dex.

\subsection{Cluster Abundances}

In order to ensure that we are working with a self-consistent set of abundances for the field dwarf
sample and the cluster sample, we adopted for reference the abundance scale of G00 for the clusters
in this study; as previously discussed, the Hyades [Fe/H] listed by G00 agrees well
with the result from our Str\"{o}mgren calibration for the field dwarfs.
Table~\ref{FEdata} 
lists the cluster abundances from G00, along with cluster ages taken from Pinsonneault et al. 
(\cite{pinson2}).  

\begin{table}
\centering
  \caption[]{Reference Metallicities and Ages of Open Clusters}
  \begin{tabular}{lcc} \hline 
            Cluster
            & ${\rm [Fe/H]_{G00}}$
            & Age (Myr) \\
            \hline

  Hyades   & +0.13$\pm$0.06 & 650 \\
  Praesepe & +0.04$\pm$0.06 & 600 \\
  Pleiades & $-$0.03$\pm$0.06 & 100 \\
  Coma Ber & $-$0.05$\pm$0.06 & 500 \\
  NGC 2516 & $-$0.16$\pm$0.11 & 140 \\
\hline 				
\label{FEdata} 
\end{tabular} 
\end{table}


\section{Empirical MS-Fitting Technique}

The basic MS-fitting method consists of using the field dwarf sample defined above to construct a
template main sequence for comparison with the main sequence of the cluster in question.  This
necessitates `shifting' each sample star in colour to account for the differences in metallicity between
the individual stars (which all have different abundances, within the specified range) and the 
cluster.  Then the shift in $V$-magnitude required to match the template MS to the cluster main 
line (after cluster reddening is accounted for) represents the distance modulus, $(m-M)_{0}$.
 
We note here that, to be physically correct, the procedure of shifting the field dwarfs to match the 
metallicity of a particular cluster should preserve mass (Reid \cite{reid}).  This implies that 
adjustments to both magnitude and colour would be necessary to match each field dwarf to a star of 
equivalent mass, at the metallicity of the cluster.  However, since mass also evolves along
the MS, at a given colour, luminosity and metallicity, the mass of a star also depends on its age.
Since in general we do not have reliable ages for local stars it is not usually possible to apply
these magnitude shifts in a physically correct way.  In Paper I, we tested the effect of applying 
colour and magnitude corrections to the subdwarf sample (for which $-1.0 < {\rm [Fe/H]} < -0.3$) assuming
an arbitrary fixed age of 10~Gyr.  
The resulting template MS covered a different range of magnitudes than the one which included colour 
shifts only, as shifting a star of fixed mass to a lower metallicity increases its luminosity, and 
vice-versa.  However, since the shape of the isochrones is essentially the same across the metallicity 
range we are using (see Sect. 4.1), the shape of the resulting template is not changed and, consequently, 
fits to these revised templates yield the same distance moduli as the templates constructed using colour 
shifts only.

\begin{figure}
     \resizebox{\hsize}{!}{\includegraphics{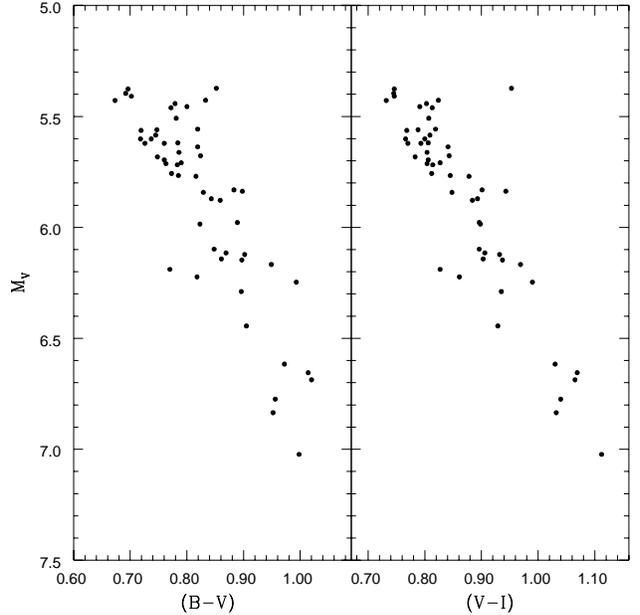}}
      \caption{CMDs for the full sample of 54 field dwarfs
              }
         \label{dwarfdata}
   \end{figure}

Colour-magnitude diagrams for the field dwarf sample, before any shifts have been applied, are 
displayed in Fig.~\ref{dwarfdata}.  The $(B-V)$ and $(V-I)$ colours are the observed values, as listed in 
Table~\ref{SDdata}, and the $M_{\rm V}$ values are the LK-corrected absolute magnitudes.

Since one of the aims of this work is to explore whether discrepancies between \emph{Hipparcos}
parallax distances and MS-fitting distances for some clusters are due to model dependencies in
MS-fitting methods, we want to derive the metallicity dependent colour shifts empirically i.e.
we do not want to rely on model isochrones at any stage in the procedure, so that our results 
will not be model dependent.


\subsection{Method}

We make only one assumption, which is that the \emph{shape} of the isochrones is the same for the
portion of the lower MS that we use in the fits, across the narrow metallicity range we have 
defined.  All the field dwarfs which are used to construct the template MS lie in the absolute magnitude
range $5.4 \leq M_{\rm V} \leq 7.0$, where $M_{\rm V}$ is the LK corrected magnitude, and the metallicity
range is as defined by the subset of the FMS97 sample used in the metallicity calibration, namely
$-0.45 \leq {\rm [Fe/H]} \leq 0.35$.  The theoretical models of Girardi et al. (\cite{girardi}) show that
the isochrones should be parallel in this range in both colour planes.  Also, if we combine the 
local field dwarfs used in this study with those in Paper I (which are in the range 
$-1.0 < {\rm [Fe/H]} < -0.3$), there is no measurable change in slope when the data are binned by 
metallicity, in either colour plane.

Using this assumption of similarity of shape for isochrones of different metallicities, the 
metallicity dependent colour shifts are calculated by first determining the colour that each
field dwarf \emph{would} have at a fixed absolute magnitude of $M_{\rm V}=6$, which we label 
$(B-V)_{M_{\rm V}=6}$ or $(V-I)_{M_{\rm V}=6}$.  We then look at the relationship between 
$(B-V)_{M_{\rm V}=6}$
(or $(V-I)_{M_{\rm V}=6}$) and metallicity, [Fe/H], to calculate $\Delta (B-V)$ (or $\Delta (V-I)$)
as a function of $\Delta {\rm [Fe/H]}$.
The observed colours of the field dwarfs were `corrected' to their appropriate values at $M_{\rm V}=6$ 
using the slope of the Hyades main sequence, effectively as an empirical isochrone.  The Hyades slope was 
derived using only single stars as defined by Perryman et al. (\cite{perry}), at their absolute 
$V$-magnitudes, as determined from the improved \emph{Hipparcos} parallaxes of Madsen et al. 
(\cite{madsen}).  Fits were made in the $V/(B-V)$ and $V/(V-I)$ planes, using the $V_{\rm J}$ 
magnitudes and $(B-V)_{\rm J}$ and $(V-I)_{\rm C}$ colours as listed in the \emph{Hipparcos} catalogue 
(fields H5, H37 and H40 respectively). 
 
The derived metallicity dependencies were:  
\newline
$\Delta(B-V) = 0.154 \Delta {\rm [Fe/H]}$ and 
$\Delta(V-I) = 0.103 \Delta {\rm [Fe/H]}$.  It is important to note that neither the assumed metallicity
nor the parallax distance of the Hyades play a part in this process, we are merely using the 
\emph{shape} of the Hyades MS.  In fact, the derived dependencies are rather insensitive to the
exact slope assumed for the main sequence.  In the procedure employed above, the Hyades MS was
fitted to a cubic function, but the relevant portion in the $V/(B-V)$ plane is actually almost 
linear, with a gradient of $\sim$5.
As a quick check, we tested the effect of assuming a MS gradient of 8, a value far outside the 
expected range of MS slopes.  With this assumed slope, the metallicity dependence in $(B-V)$ 
becomes $\Delta(B-V) = 0.132 \Delta {\rm [Fe/H]}$, and the change in the distance moduli we ultimately 
derive is no more than 0.03 mag.

Using the above relationships, the appropriate colour shifts 
are applied to each field dwarf in the sample, at their absolute magnitudes, to build up a template MS
at the metallicity of each cluster to be studied.  The template MS corresponding to each
particular cluster is then shifted in $V$-magnitude to match the main line of the cluster in 
question, the shift in $V$ being equal to the distance modulus, $m-M$, and the best-fit to the 
cluster main line is found by using a least-squares fitting routine.  Weighted errors are used
for the field dwarfs and include both photometry errors and magnitude errors due to errors on the 
parallax, where $\Delta M_{\rm V} = 2.17(\Delta \pi/\pi)$.  Errors on the metallicities of both the
field dwarfs and the cluster are also accounted for and the best-fit distance modulus is found by 
minimizing $\chi^{2}$.

We note here that HIP 84164 is an obvious outlier in the CMD and may either be evolved or an 
unidentified binary.  A further 7 stars in the sample have predicted metallicities which fall 
outside the range used in our calibration.  These 8 stars have therefore been omitted from all our 
subsequent analyses, leaving a total sample of 46 field dwarfs.
Details of the fits for each cluster are presented in the next section. 


\section{Cluster by Cluster Analysis}

Main lines for each of the open clusters were derived from archive data, as detailed below, 
by fitting to a polynomial using a least-squares fitting routine and minimizing $\chi^{2}$.  
Reddening values were taken from the literature and the cluster metallicities of G00 were used, 
unless otherwise stated.

\subsection{Hyades}

$V_{\rm J}$, $(B-V)_{\rm J}$ and $(V-I)_{\rm C}$ data for the Hyades were taken directly from the 
\emph{Hipparcos} catalogue (fields H5, H37 and H40, respectively).  Most of the data for the 
Hyades listed in these fields are taken from ground-based photoelectric observations, 
and a few of the $(V-I)_{\rm C}$ values are transformations from other indices e.g. $(V-I)_{\rm J}$ or 
$(R-I)_{\rm C}$.
We restricted the data to definite cluster members listed as single and non-variable in Perryman 
et al. (\cite{perry}) and made an initial cut of $V_{\rm J} > 6.0$, as we are only 
interested in the lower MS.  In order to make an unbiased estimate of the Hyades distance, we must
derive the main line from the \emph{apparent} magnitudes of the MS stars.  Since the Hyades is so
close, a CMD of all the stars in this subset shows quite a large spread in apparent magnitude, as 
the cluster structure is effectively resolved.  Hence, we restricted the sample to stars lying 
within 10 pc of the cluster centre, as defined by Perryman et al. (\cite{perry}), which is the 
same cut made in that study to derive the average \emph{Hipparcos} parallax distance to the 
cluster.  A final cut in colour was made at $0.4 \leq (B-V) \leq 1.4$, which resulted in a sample 
of 58 Hyads.  The same subset of stars was used to fit in the $V/(V-I)_{\rm C}$ plane, resulting in a 
colour range of $0.45 < (V-I)_{\rm C} < 1.55$.  We assumed zero reddening for the cluster and the G00
metallicity value of ${\rm [Fe/H]} = 0.13 \pm 0.06$, which is consistent with our determination of 
${\rm [Fe/H]} = 0.122$.  

Shifting the whole field dwarf sample to ${\rm [Fe/H]} = 0.13$ and fitting to the derived cluster main 
lines results in distance moduli of $(m-M)_{0} = 3.33 \pm 0.06$ from the $(B-V)$ index and
$(m-M)_{0} = 3.32 \pm 0.04$ from $(V-I)_{\rm C}$.  Thus we reproduce precisely the \emph{Hipparcos} 
parallax distance modulus of $(m-M)_{0} = 3.33 \pm 0.01$ (Perryman et al.~\cite{perry}).

\begin{figure}
     \resizebox{\hsize}{!}{\includegraphics{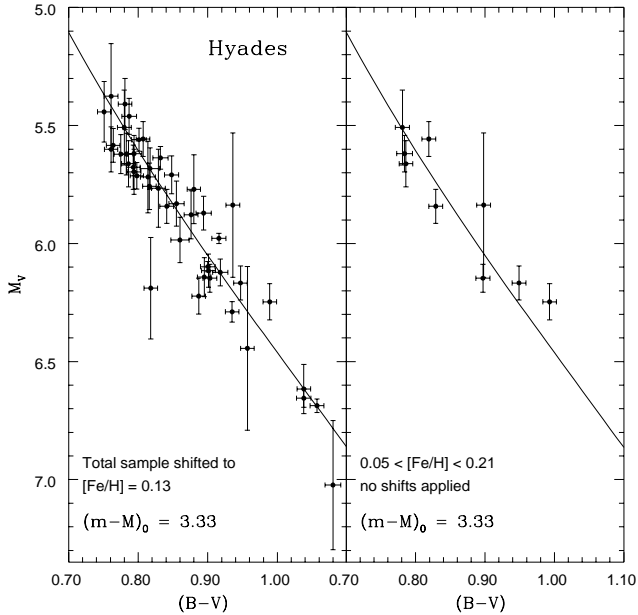}}
      \caption{Hyades fits in $V/(B-V)$.  The data points represent the field dwarfs at the cluster 
metallicity, with error bars as indicated in Sect. 4.1.  The solid line in each panel is the cluster 
main-line shifted in magnitude to the best fit distance modulus, as indicated (see text for details).
              }
         \label{hyfit_bv}
   \end{figure}

\begin{figure}
     \resizebox{\hsize}{!}{\includegraphics{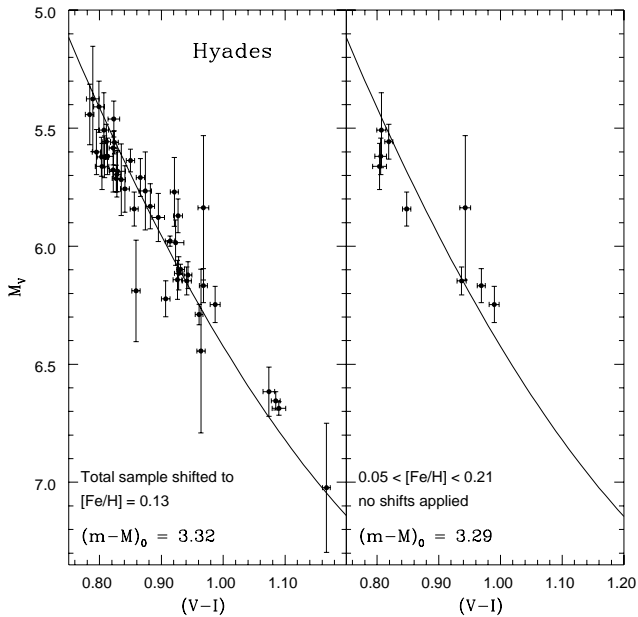}}
      \caption{Hyades fits in $V/(V-I)_{\rm C}$.  Comments as for Figure 2. 
              }
         \label{hyfit_vi}
   \end{figure}

As a consistency check, we performed the fits using the subset of 9 stars with [Fe/H] values 
within $\pm$ 0.08 of the Hyades metallicity (with an average of ${\rm [Fe/H]} = 0.122$, coincidentally).
Since these stars have, within the errors, the same metallicity as the cluster, no shifts in colour
were applied.  The resulting distance moduli were:
$(m-M)_{0} = 3.33 \pm 0.09$ from $(B-V)$ and
$(m-M)_{0} = 3.29 \pm 0.07$ from $(V-I)_{\rm C}$, thus we are sure there are no systematic effects 
arising from our derived metallicity dependencies. 
Figs.~\ref{hyfit_bv} and \ref{hyfit_vi} show the Hyades main line shifted to its best-fit distance
in, respectively, the $V/(B-V)$ and $V/(V-I)_{\rm C}$ planes.  In each case, the fit to the full `shifted'
field dwarf sample is shown in the left panel, whilst the fit to the `unshifted' subset is shown on the 
right.
\subsection{Praesepe}

Data for Praesepe were taken from the WEBDA data base at http://obswww.unige.ch/webda/.  All the
available $(B-V)$ and $(V-I)$ photoelectric data were used to minimize the effect of systematic 
offsets between the various data sets and averages were taken for stars with multiple observations.
The $(B-V)_{\rm J}$ data are principally made up of samples taken from Johnson (\cite{johnson}), Mendoza
(\cite{mendoza}), Upgren et al. (\cite{upgren}), Weis (\cite{weis}), Stauffer (\cite{stauffer}) and
Mermilliod et al. (\cite{merm}).  All the archive $(V-I)$ data for Praesepe are 
either on the Johnson or the Kron systems, rather than Cousins.  $(V-I)_{\rm J}$ data are from Mendoza
(\cite{mendoza}) whilst the $(V-I)_{\rm K}$ data are made up of the samples from Upgren 
(\cite{upgren}), Weis (\cite{weis}) and Stauffer(\cite{stauffer}).  The $(V-I)$ data were converted
to the $(V-I)_{\rm C}$ system using the transformations of Bessell (\cite{bessell}) and Bessell 
\& Weis (\cite{besweis}), where $(V-I)_{\rm C} = 0.778(V-I)_{\rm J}$ and 
$(V-I)_{\rm C} = 0.227 + 0.9567(V-I)_{\rm K} + 0.0128(V-I)_{\rm K}^{2} -0.0053(V-I)_{\rm K}^{3}$.
Known binaries from Bouvier et al. (\cite{bouvier2}) were removed and initial cuts were made at 
$0.4 < (B-V) < 1.4$ and $0.45 < (V-I)_{\rm C} < 1.8$.  The cluster main lines were found by successively 
fitting to a function, making 1$\sigma$ cuts and then re-fitting, until the solutions converged. 
This effectively removes any field contamination and unidentified binaries from the sample. 

\begin{figure}
     \resizebox{\hsize}{!}{\includegraphics{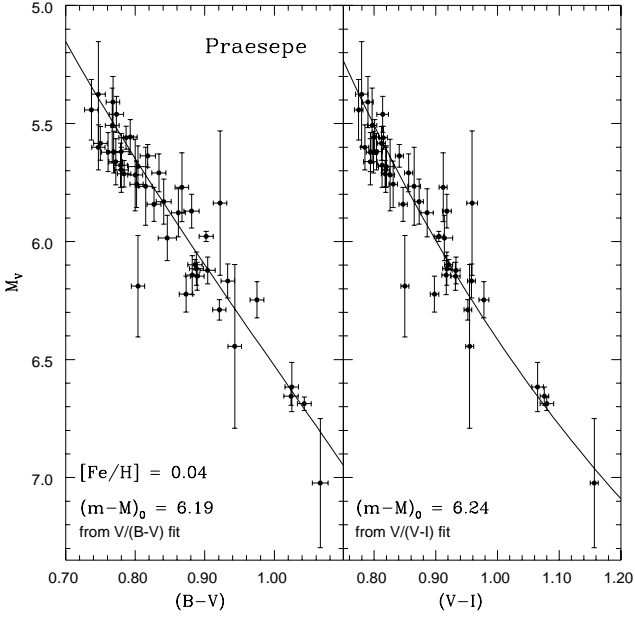}}
      \caption{Praesepe fits in $V/(B-V)$ and $V/(V-I)_{\rm C}$
              }
         \label{prfit}
   \end{figure}

Assuming zero cluster reddening and shifting the field dwarf sample to the G00 metallicity value of 
${\rm [Fe/H]} = 0.04 \pm 0.06$ results in distance moduli of $(m-M)_{0} = 6.19 \pm 0.06$ from the 
$(B-V)$ fits and $(m-M)_{0} = 6.24 \pm 0.04$ from $(V-I)_{\rm C}$.  These results are just consistent,
within the 1$\sigma$ errors, with the \emph{Hipparcos} parallax
distance found by 
van Leeuwen~(\cite{vanl}) of $(m-M)_{0} = 6.37 \pm 0.15$.  A recent
study of cluster reddenings by Taylor \& 
Joner (\cite{tayjon}) has found an average reddening of $E(B-V) = 0.022$ for Praesepe.  If this 
reddening is assumed, the cluster distance modulus increases by $\sim$ 0.05 mag, making it more 
consistent with the \emph{Hipparcos} result.
Fig.~\ref{prfit} shows the best-fits to the Praesepe main line in $V/(B-V)$ (left panel) and
$V/(V-I)$ (right panel) assuming ${\rm [Fe/H]} = 0.04$ and zero reddening.

\subsection{Pleiades}

Photoelectric data for the Pleiades were also taken from WEBDA and treated in a similar manner to 
those for Praesepe.  $(B-V)_{\rm J}$ data are principally from Johnson \& Mitchell (\cite{johnmit}),  
Mendoza (\cite{mendoza}), Landolt (\cite{landolt}) and Stauffer (\cite{stauffer2}).  The available 
$(V-I)$ data have been transformed onto the Cousins system (see Sect. 5.2) -- the $(V-I)_{\rm J}$ sample 
being from Mendoza (\cite{mendoza}) and $(V-I)_{\rm K}$ principally from Stauffer (\cite{stauffer2}).
Averages were taken for stars with multiple observations and initial cuts made at 
$0.4 < (B-V) < 1.4$ and $0.45 < (V-I)_{\rm C} < 1.8$.  Known binaries from Bouvier et al. 
(\cite{bouvier1}) and Raboud \& Mermilliod (\cite{raboud}) were removed from the samples.
The Pleiades is known to have patchy reddening and so individual stars were dereddened and 
extinction corrected using the values from Breger (\cite{breger}).  For stars with no reddening 
listed by Breger, an average value of $E(B-V) = 0.04$ was assumed.  The main lines were derived 
as for Praesepe, by making successive fits and 1$\sigma$ cuts, until the solution converged.

\begin{figure}
     \resizebox{\hsize}{!}{\includegraphics{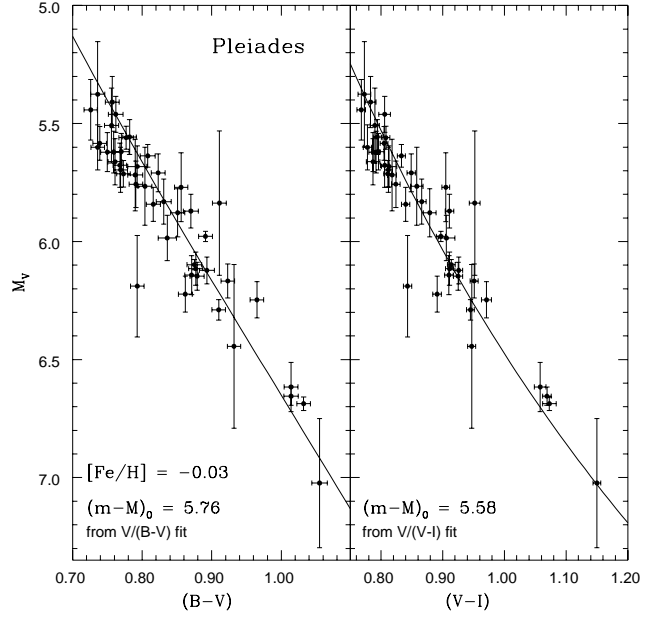}}
      \caption{Pleiades fits in $V/(B-V)$ and $V/(V-I)_{\rm C}$
              }
         \label{plfit1}
   \end{figure}

Using the G00 metallicity value of ${\rm [Fe/H]} = -0.03 \pm 0.06$ leads to a dereddened distance 
modulus of $(m-M)_{0} = 5.76 \pm 0.06$ from the $(B-V)$ fits and $(m-M)_{0} = 5.58 \pm 0.04$ from 
$(V-I)_{\rm C}$ (see Fig.~\ref{plfit1}). These two distance moduli are
mutually inconsistent and significantly
longer than the \emph{Hipparcos} parallax distance of
$(m-M)_{0} = 5.37 \pm 0.07$ (van Leeuwen \cite{vanl}).

Performing fits using just the subset of stars with abundances within $\pm$ 0.1 dex of the assumed
cluster metallicity, and not applying any colour shifts, produces exactly the same distance moduli
in both colour planes and so there do not appear to be any systematic effects arising from our 
derived metallicity dependencies.
We also note here that our dereddened Pleiades main line in $V/(B-V)$ exactly matches that derived by 
Turner (\cite{turner}), who, assuming solar abundance for the Pleiades, finds a distance 
modulus of 5.56, assuming 3.15 for the Hyades i.e. a difference of 2.41 mag.  This compares well with
our difference of 2.43 mag from these fits at ${\rm [Fe/H]} = -0.03$.


\subsection{NGC 2516}

\begin{figure}
     \resizebox{\hsize}{!}{\includegraphics{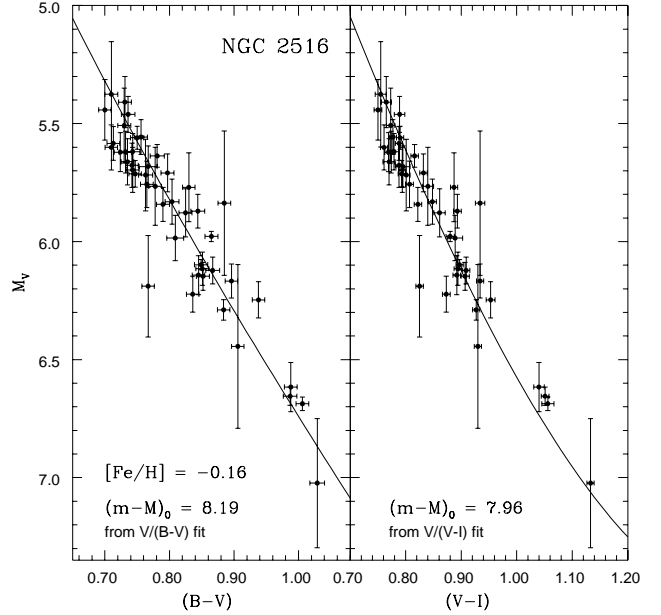}}
      \caption{NGC 2516 fits in $V/(B-V)$ and $V/(V-I)_{\rm C}$
              }
         \label{2516fit}
   \end{figure}

For NGC 2516 we used $(B-V)_{\rm J}$ and $(V-I)_{\rm C}$ CCD data from the study by Jeffries, Thurston \& 
Hambly (\cite{jeff}).  Jeffries et al. (\cite{jeff}) provide membership information on the basis 
of a star's position in the $V/(B-V)$, $V/(V-I)_{\rm C}$ and $(B-V)/(V-I)$ planes, and likely binaries 
are also flagged.  We derived main lines from those stars listed as definite members, after 
removing those flagged as binaries, using the same procedures as for the other clusters.  Assuming
a reddening of $E(B-V) = 0.12$ (Dachs \& Kabus \cite{dachs}) and the G00 listed metallicity of 
${\rm [Fe/H]} = -0.16 \pm 0.11$ the best-fit distance moduli are: 
$(m-M)_{0} = 8.19 \pm 0.10$ from $(B-V)$ and $(m-M)_{0} = 7.96 \pm 0.07$ from $(V-I)_{\rm C}$ 
(see Fig.~\ref{2516fit}).
It is immediately apparent that a situation arises which is similar to that for the Pleiades in that 
the MS-fitting distances are discrepant between the 2 colour indices when the G00 metallicity is 
adopted, and both are longer than the \emph{Hipparcos} parallax distance of 
$(m-M)_{0} = 7.70^{+0.16}_{-0.15}$ (Robichon et al. \cite{robi}).


\section{Discussion}

\begin{figure}
     \resizebox{\hsize}{!}{\includegraphics{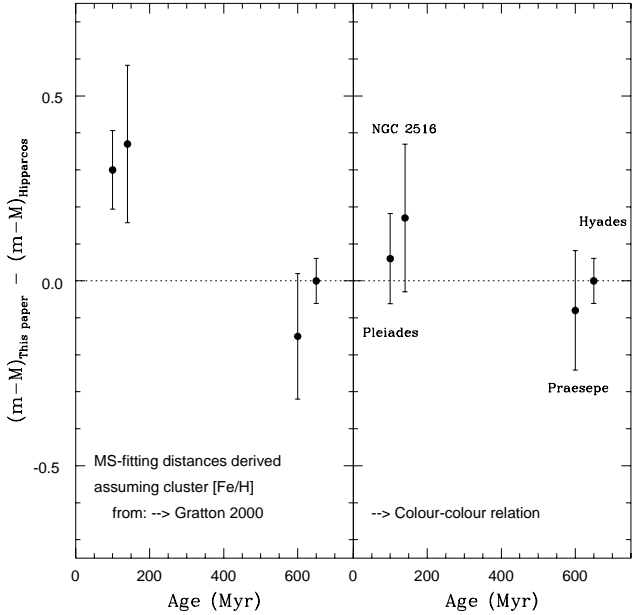}}
      \caption{Relationship between cluster age and $\Delta (m-M)^{\rm MSF}_{\rm Hip}$
              }
         \label{age_dist}
   \end{figure}

Employing our empirical MS-fitting method, we derive distance moduli for the Hyades and Praesepe 
which are in agreement with the \emph{Hipparcos} parallax distances, when using the cluster 
spectroscopic metallicity scale by G00. However, for both the Pleiades and NGC~2516 the distance
moduli we find are discrepant with respect to those from \emph{Hipparcos}, in qualitative agreement 
with the findings by P98 (in the case of the Pleiades) and Terndrup et al. (\cite{tern}, in the case 
of NGC~2516).
These results are shown in the left panel of Fig.~\ref{age_dist}, where we display the difference 
between our derived distance moduli and those from \emph{Hipparcos}, $\Delta (m-M)^{\rm MSF}_{\rm Hip}$ 
(where $\Delta (m-M)^{\rm MSF}_{\rm Hip} = (m-M)_{\rm This~paper} - (m-M)_{\rm Hipparcos}$) against the 
age of the cluster (taken from Pinsonneault et al. \cite{pinson2}, see table~\ref{FEdata}).  Note that in
the case of the Pleiades and NGC~2516 we have taken the average of the distance moduli in $V/(B-V)$ and 
$V/(V-I)_{\rm C}$.  The error bars represent the 1$\sigma$ errors from this paper and from van Leeuwen 
(\cite{vanl}), added in quadrature.  

These discrepancies in the Pleiades and NGC~2516 distances have already been interpreted as an 
indication of systematic errors in the \emph{Hipparcos} parallaxes for these two clusters; these 
(hypothetical) errors would amount to $\sim 1-2$ mas (P98) in case of the Pleiades, and only 
$\sim$0.3 mas in case of NGC~2516 (Terndrup et al. \cite{tern}).  Systematic errors in the 
\emph{Hipparcos} parallaxes have however been ruled out by van Leeuwen (\cite{vanl}) and more recently
by Makarov (\cite{makarov}).  

There is another possible explanation for this mismatch, related to the possibility that the colours of 
the Pleiades and NGC~2516 MS stars are too blue for their nominal spectroscopic [Fe/H].
When applying the MS-fitting technique, the field dwarf colours have in principle to be shifted to the
intrinsic colours of the cluster MS, so that the vertical shift necessary to superimpose the two 
sequences is a true measure of the cluster distance modulus. It is therefore paramount that the 
spectroscopic [Fe/H] scale for the clusters is consistent with both the subdwarf and cluster 
colours.  We are in a position to test this concurrence by using the $(B-V)-(V-I)_{\rm C}$ colour-colour 
diagrams. Once the cluster reddening is accounted for, the position of a cluster main line (or 
field dwarf sequence) in this colour-colour diagram is a function of its metallicity, and can be 
compared with the sequence of subdwarfs shifted to the nominal cluster spectroscopic metallicity.

\begin{figure}
     \resizebox{\hsize}{!}{\includegraphics{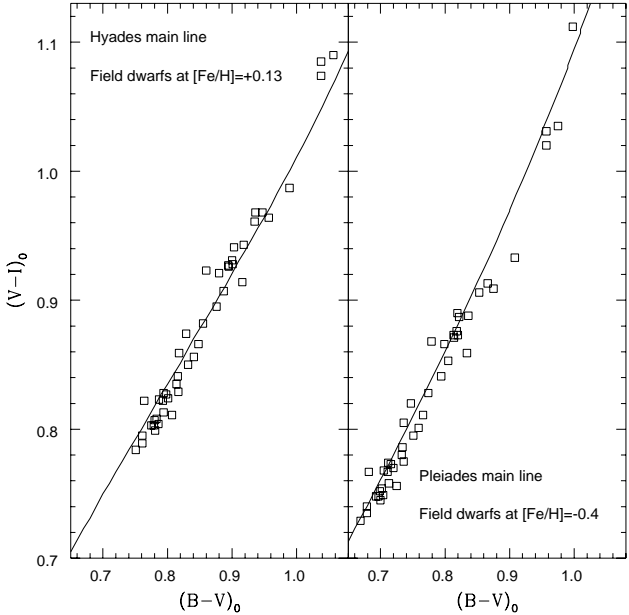}}
      \caption{Colour-colour plots for the Hyades, Pleiades and field dwarf sample
              }
         \label{colcol}
   \end{figure}

\begin{figure}
     \resizebox{\hsize}{!}{\includegraphics{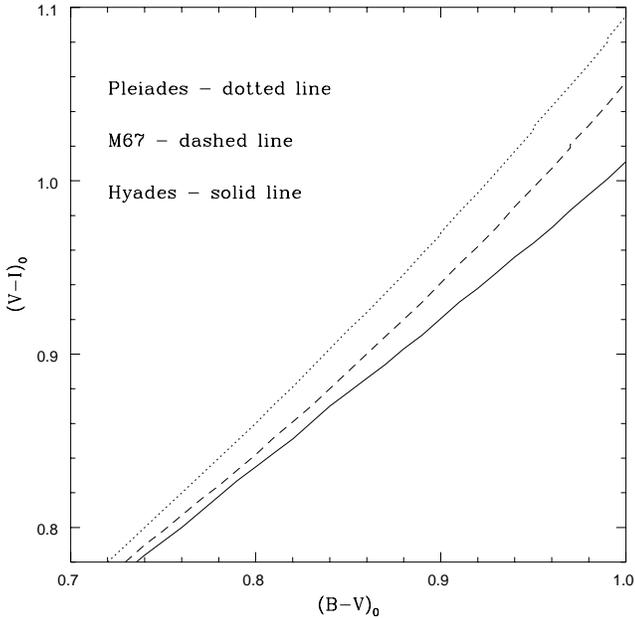}}
      \caption{Colour-colour plots for the Pleiades, M67 and Hyades
              }
         \label{cols2}
   \end{figure}

In Fig.~\ref{colcol} we show, as an example, the dereddened colours for the Hyades and Pleiades MS 
plotted together with the colours of the whole field dwarf sample shifted in metallicity to 
${\rm [Fe/H]} = 0.13$ and ${\rm [Fe/H]} = -0.4$, respectively. 
Whereas the Hyades line nicely agrees with the field dwarf colours at the Hyades metallicity, the 
Pleiades colours are consistent with the cluster abundance being significantly subsolar at 
${\rm [Fe/H]} \sim -0.4$, rather than the near solar abundance indicated from spectroscopy.
For reference, we show in Fig.~\ref{cols2} the Hyades and Pleiades colour-colour relations, as 
plotted in Fig.~\ref{colcol}, along with the main line for the old open cluster M67 which has 
approximately solar abundance from high resolution spectroscopy (G00 gives ${\rm [Fe/H]} = +0.02$).  
Data for M67 are from Montgomery et al. (\cite{mont}) and the reddening value of $E(B-V)=0.04$ is taken 
from Twarog et al. (\cite{twar97}).

This mismatch between the Pleiades spectroscopic and photometric metallicity fully explains the 
discrepancy between the MS-fitting distance moduli in  $V/(B-V)$ and $V/(V-I)_{\rm C}$, and with the 
\emph{Hipparcos} distance.  In fact, assuming a value ${\rm [Fe/H]} = -0.4 \pm 0.1$ appropriate to the 
Pleiades intrinsic colours and rederiving the MS-fitting distance, we obtain
$(m-M)_{0} = 5.46 \pm 0.09$ from the $(B-V)$ colour and $(m-M)_{0} = 5.39 \pm 0.06$ from 
$(V-I)_{\rm C}$. These results are now mutually consistent and also in agreement with the 
\emph{Hipparcos} parallax distance of $(m-M)_{0} = 5.37 \pm 0.07$ (van Leeuwen \cite{vanl}).

We repeated the same analysis for NGC~2516 and Praesepe.  In the first case we obtain a 
metallicity slightly lower than the Pleiades one, that is ${\rm [Fe/H]} = -0.5$, which again provides 
consistent distances.  If we adopt ${\rm [Fe/H]} = -0.5 \pm 0.1$ and redo the fits, the best-fit distance 
moduli are:  
$(m-M)_{0} = 7.95 \pm 0.10$ from $(B-V)$ and $(m-M)_{0} = 7.79 \pm 0.07$ from $(V-I)_{\rm C}$.   
These results are also in agreement, within their 1$\sigma$ errors, with the \emph{Hipparcos} 
parallax distance of $(m-M)_{0} = 7.70^{+0.16}_{-0.15}$ (Robichon et al. \cite{robi}).

In the case of Praesepe the photometric metallicity happens to be the same as that of the Hyades, a 
result also in agreement with the spectroscopic analysis by Boesgaard et al. (\cite{boesbu}), and 
in any case not dramatically different from the G00 result.  At the Hyades metallicity the derived 
distance moduli become: 
$(m-M)_{0} = 6.26 \pm 0.06$ from $(B-V)$ and $(m-M)_{0} = 6.28 \pm 0.04$ from $(V-I)_{\rm C}$ 
(assuming a zero reddening; they are increased by 0.05 mag if we use E$(B-V)$=0.022). Both results 
are completely consistent with the \emph{Hipparcos} distance $(m-M)_{0} = 6.37 \pm 0.15$ 
(van Leeuwen~\cite{vanl}).

The right panel of Fig.~\ref{age_dist} summarizes these results obtained from the photometric 
metallicities; the previous discrepancies with respect to the \emph{Hipparcos} parallaxes have now 
disappeared.  We conclude that our analysis does not support any discrepancy between MS-fitting and 
\emph{Hipparcos} distances, at least for the clusters we have studied. The key point is the 
realization that the colours of the MS of the Pleiades and NGC~2516 are not appropriate for their 
spectroscopic metallicities. It is important to stress that this occurrence is not invoked as an 
hypothetical explanation for the distance mismatch obtained using the spectroscopic metallicities, 
but it is demonstrated by comparing the cluster and subdwarf colour-colour diagrams. 
A MS-fitting based distance is meaningful only if the colours of the reference MS are consistent with
the cluster ones, and this is clearly not the case for the Pleiades and NGC~2516 when the G00 
[Fe/H] values are used.  It is the photometric metallicity based on the colour indices used in the 
MS-fitting procedure that has to be used in order to get a meaningful MS-fitting distance, not the 
spectroscopic one, if this is different. 
Therefore the widely discussed discrepancy between MS-fitting and \emph{Hipparcos} distances for 
these two clusters is just an artifact due to the clusters' colours which are inconsistent with their
spectroscopic metallicity.

A possible explanation for this colour mismatch would be that the field dwarf [Fe/H] scale (and
hence colours) is inconsistent with the cluster one. However, we have already seen that the field dwarf 
[Fe/H] scale predicts the G00 Hyades metallicity, which is also confirmed by many independent 
spectroscopic analyses.  It is therefore not possible to claim a zero point offset between our 
field dwarf metallicity scale and that of G00. The case of Praesepe, whose metallicity based on the MS 
colours is only slightly different from the G00 one and in agreement with independent spectroscopic
estimates, further confirms the absence of any substantial offset.

Another possibility is a much higher helium content for the Pleiades and NGC~2516 than for the local 
field dwarfs.  P98 and Stello \& Nissen (\cite{stello}) discuss the possibility that anomalous results 
for the Pleiades may be caused by a higher than expected He abundance.  Both authors conclude
that this is unlikely because the over-abundance required would be very large, although this
cannot be entirely ruled out without further spectroscopic studies of both the Pleiades and
other young clusters.

It is not easy to find an explanation for this discrepancy between spectroscopic and photometric 
[Fe/H] for Pleiades and NGC~2516. Unless a combination of errors in the spectroscopic metallicity, 
photometry and reddening conspire to produce the observed mismatch, one can only speculate on the 
origin of this problem. It is interesting to note that it affects the two younger clusters which 
should contain faster rotators and have higher chromospheric activity levels on average. 
P98 suggest that the colour-temperature relation may be different for
fast and slow rotators -- their Figure~10 appears to show that Pleiades fast rotators lie below the
cluster main line in the $V(B-V)$ CMD for colours redder than $(B-V) \sim 0.9$, but that the same stars 
lie \emph{on} the main line in $V/(V-I)$.  Also NGC~2516 members may have a pattern of rotation rates
similar to the Pleiades stars (see Terndrup \cite{tern}, their Figure~5).  

It should be noted however, that predictions from theory do not necessarily agree with the trend 
suggested by the empirical data.  For solar metallicity stars with masses greater than 1.5 $M_{\odot}$, 
Hern\'{a}ndez et al. (\cite{hern}) find that the predicted colours are redder for the faster rotators, 
whilst for low metallicity stars ([Fe/H] $< \sim -1.0$) on the MS and around the turnoff, Deliyannis et 
al. (\cite{deli}) find no significant difference between their standard and rotational models. 
Evidently, this disparity between theory and observation requires a more detailed investigation, 
which is well beyond the scope of the present work.  It is clear however, that rotation may well be a
factor contributing to the apparently anomalous colours in young clusters, which could potentially lead 
to a discrepancy between the spectroscopic and photometric abundance estimates.

Sung et al.~(\cite{sung}) have discussed the possibility that chromospheric activity produces a blue 
excess in NGC~2516 G- and K-star colours, an occurrence dismissed by Terndrup et al.~(\cite{tern}), 
who however find blue excesses for late-K dwarfs in the Pleiades.  In a very recent study of active 
binary stars, Katz et al. (\cite{katz}) find that, when deriving effective temperatures from colour 
indices for these active stars, the $(V-I)$ index gives significantly cooler estimates than the 
$(B-V)$ index.  This offset is in the same sense that we find for the 2 young clusters i.e. at 
fixed $(B-V)$, the $(V-I)$ colours are too red for their spectroscopic abundance.
In an attempt to further test this possibility we have considered the chromospheric activity study 
by Henry et al.~(\cite{henry}), and selected stars with either very low or very high activity, in 
order to test if their $(B-V)-(V-I)$ colour-colour relationship is a function of the level of
chromospheric activity. Unfortunately the low number of objects with both colours available 
prevented us from drawing any meaningful conclusions on this issue, which in our opinion needs a 
deeper analysis.


\begin{acknowledgements}
We are very grateful to Francois van Wyk at the SAAO for observations of the brightest stars in our 
sample, taken with the 0.5m telescope.
It is a pleasure to thank Phil James for many useful discussions and helpful suggestions which have 
improved this paper.  We also thank the anonymous referee for an insightful reading of the manuscript 
and some pertinent suggestions.  
SMP acknowledges financial support from PPARC. 
This research has made extensive use of the \emph{Hipparcos}, SIMBAD and WEBDA databases.

\end{acknowledgements}

\end{document}